\documentclass{elsart}
\usepackage{harvard}

\def\oiii{[{\sc O\, iii}]}
\def\oii{[{\sc O\, ii}]}
\def\feka{Fe K$\alpha$}
\def\ltsima{$\; \buildrel < \over \sim \;$}
\def\simlt{\lower.5ex\hbox{\ltsima}} 
\def\gtsima{$\; \buildrel > \over \sim \;$}
\def\simgt{\lower.5ex\hbox{\gtsima}} 

\begin{document}
\input{psfig.tex}

\runauthor{Sambruna, Eracleous, \& Mushotzky}
\begin{frontmatter}
\title{Constraining the central engine of radio-loud AGNs}
\author[State College]{Rita M. Sambruna},
\author[State College]{Michael Eracleous},
\author[GSFC]{Richard F. Mushotzky}

\address[State College]{Department of Astronomy and Astrophysics, The 
Pennsylvania State University, 525 Davey Lab, State College, PA 16802}
\address[GSFC]{NASA/GSFC, Code 662, Greenbelt, MD 20771}

\begin{abstract}
We present an X-ray spectral survey of radio-loud (RL) AGNs using {\it
ASCA} and {\it RXTE}. The goal was to study the structure of their
central engines and compare it to that of their radio-quiet (RQ)
counterparts.  We find Systematic differences in the X-ray properties
of the two AGN classes. Specifically, RL AGNs exhibit weaker Fe lines
and reflection components than RQ AGNs, indicating smaller solid
angles subtended by the reprocessing medium to the X-ray source. The
circumnuclear environs of RL and RQ AGNs also differ: large amounts of
cold gas are detected in Broad Line Radio Galaxies and Quasars,
contrary to what is found in Seyfert galaxies of similar X-ray
luminosity. We also discuss {\it ASCA} observations of Weak Line Radio
Galaxies, a distinct subset of radio galaxies which may harbor a
low-luminosity AGN powered by an advection-dominated accretion flow.
\end{abstract}

\end{frontmatter}

\section{Introduction}
\typeout{SET RUN AUTHOR to \@runauthor}

One of the fundamental open questions in our understanding of Active
Galactic Nuclei (AGN) is the difference between radio-loud
(RL)\footnote{We exclude blazars, whose radiation properties are
dominated by beamed emission from the jet.} and radio-quiet (RQ)
objects.  In the current AGN paradigm, their power is ultimately
provided by accretion of gas from the host galaxy onto a central black
hole. While this picture holds for both RQ and RL objects
\cite{megunif,antonucci93}, it is intriguing that the latter are able
to produce powerful, collimated relativistic jets which instead are
weak or absent in their RQ counterparts. While the environments of RL
and RQ objects are different, as are the host galaxies (RL AGNs are
found exclusively in elliptical galaxies), exactly how these large
scale phenomena influence the nature of the central engine is unknown.
More intrinsic physical differences related to the properties of the
accretion flow or the central black hole is also a candidate cause of
the dichotomy between RL and RQ AGNs. In particular, RL AGNs may
harbor rapidly spinning black holes \cite{blandzna77,meier99} and/or
very large accretion disks \cite{blandpay82}, or their inner disks may
have the form of an advection-dominated (ion) torus
\cite{rees82,narayi94,narayi95}.

The X-ray spectra of AGNs may be the means by which we understand the
difference between RL and RQ sources since the X-ray emission is
associated with the hottest, innermost parts of the accretion flow.
Indeed, models for jet formations predict different accretion
structures and thus different X-ray radiation properties.  It is
therefore encouraging that a number of systematic differences exist
between the X-ray continua of RL and RQ AGNs from earlier {\it
Einstein} and {\it EXOSAT} data \cite{wilkelv87,lawson92}, with RL AGN
having somewhat flatter X-ray spectra and higher X-ray luminosities
than RQ ones.

The advent of {\it ASCA}, with its improved spectral resolution, and
of {\it RXTE}, with its wide energy coverage and large collecting
area, has opened a new chapter in the study of the X-ray properties of
AGN. {\it ASCA} observations of Seyfert 1s established that these
objects exhibit a strong (Equivalent Width EW $\sim$ 250 eV) and broad
(FWHM $\sim$ 50,000 km/s) fluorescent \feka~ line at 6.4 keV
(rest-frame), with an asymmetric red wing \cite{nandra97}. The line
profiles are consistent with an origin from the inner parts of an
accretion disk \cite{fabian91}.  At higher energies, Seyfert 1s
exhibit a ``bump'' in their spectra peaking around 20--30 keV
\cite{nandra94,weaver98}, which is attributed to Compton reflection of
the primary, power-law continuum from a cold reprocessor, most likely
the accretion disk \cite{geofabian91}.

Soft X-ray observations of AGNs can also probe the immediate black
hole environment. In 50\% of the Seyfert 1s observed with {\it ASCA},
evidence for absorption features around 0.6--0.7 keV (the range of
ionized oxygen) were observed \cite{geo98,reynolds97}. This indicates
the presence of a ``warm absorber'' along the line of sight to the
nucleus in these systems, with densities $N{\rm _H^W} \sim 10^{21-24}$
cm$^{-2}$.

In order to study the structure of their central engines, we undertook
a systematic spectral survey of RL AGN using public {\it ASCA} daa as
well as new observations with both {\it ASCA} and {\it RXTE}. More
details can be found in \citeasnoun{SEM99} and \citeasnoun{ESM99},
together with references to previously published works.

\section{The ASCA and RXTE samples}

The {\it ASCA} sample includes all lobe-dominated radio-loud AGN with
public data up to September 1998. A subdivision was performed on the
basis of the optical spectroscopic properties \cite{SEM99}, such as
the presence of broad, permitted lines in their optical spectra and
the luminosity of the \oiii~$\lambda5007$ line. The final sample,
which is not unbiased or statistically complete by any means, includes
9 Broad Line Radio Galaxies (BLRGs), 6 Quasars, (QSRs), 12 Narrow Line
Radio Galaxies (NLRGs), and 11 Radio Galaxies (RGs), of mixed FR~I and
FR~II types. The sample objects have redshifts $z$ \ltsima 0.5.

The {\it RXTE} sample \cite{ESM99} contains four nearby, X-ray bright
BLRGs (F$_{2-10~keV} \sim 1-4 \times 10^{-11}~{\rm erg~s^{-1}}$). Of
these, 3C~120 and 3C~111 are superluminal sources and the inclination
of the jet with respect to the observer is constrained from radio
observations \cite{mike98}; lower limits on the jet inclination are
obtained for the other two BLRGs, 3C~382 and Pictor~A, from their
large scale radio morphology.

\section{Results}

A power-law spectral component in the 2--10~keV band, most likely
coming from the AGN, is detected by {\it ASCA} in 90\% of the sources.
The average photon index is $\langle \Gamma_X \rangle \sim 1.7-1.9$
for all the four subclasses (in agreement with unification models).
Similar slopes are found with {\it RXTE} for the four observed BLRGs. 
The distribution of $\Gamma_X$ among BLRGs, QSRs, NLRGs, and RGs is
not demonstrably different from RQ AGNs of matching intrinsic X-ray
luminosity. Thus, our results support previous claims that the flatter
spectra of RL AGNs are a result of a contribution from a beamed
component from the jet. This conclusion is bolstered by the fact that
no correlation between the nuclear X-ray luminosity and the core radio
luminosity is observed for our sample of (lobe-dominated) RL sources.

\begin{figure}
\noindent
{\psfig{figure=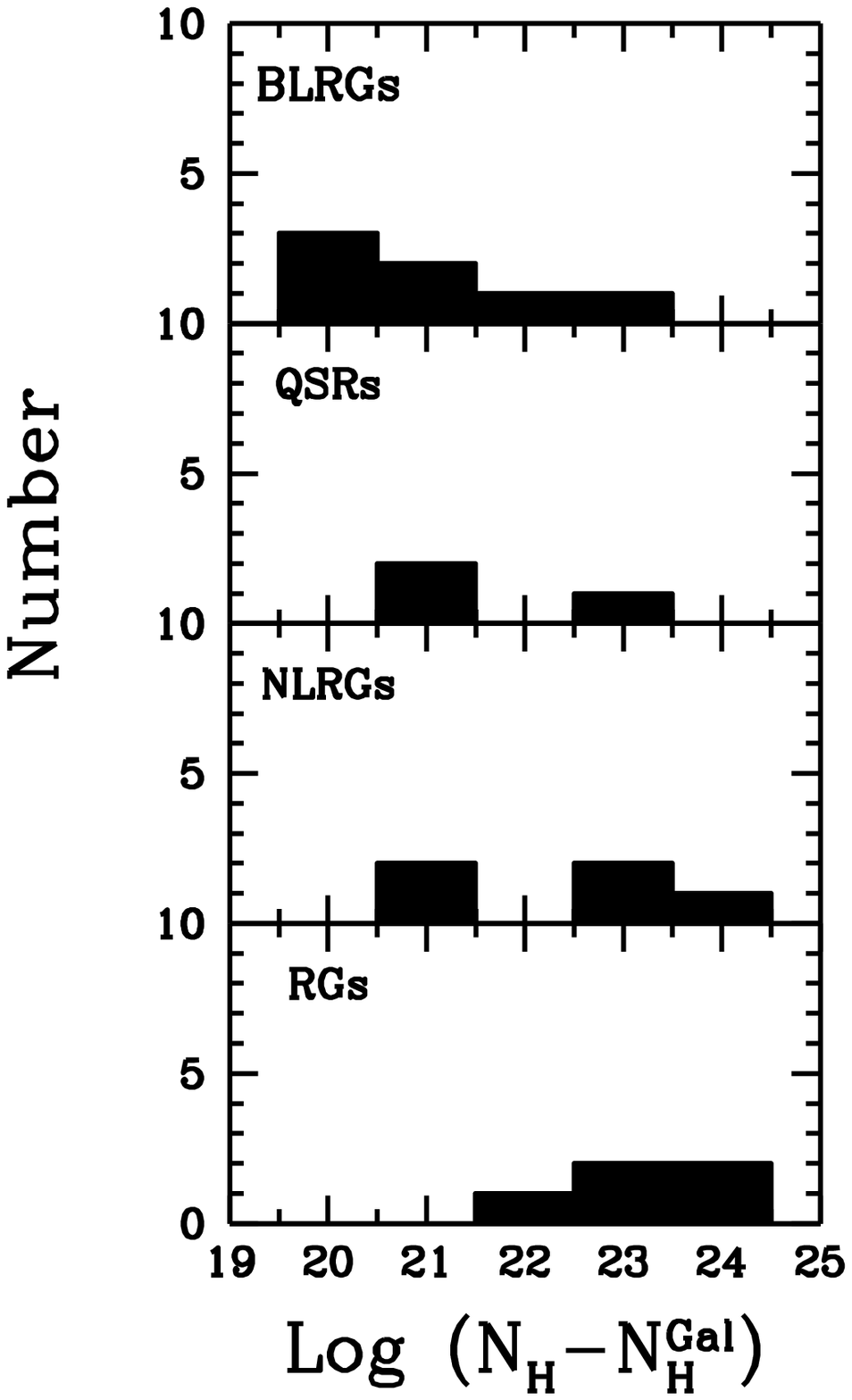,height=3.0in}}{\psfig{figure=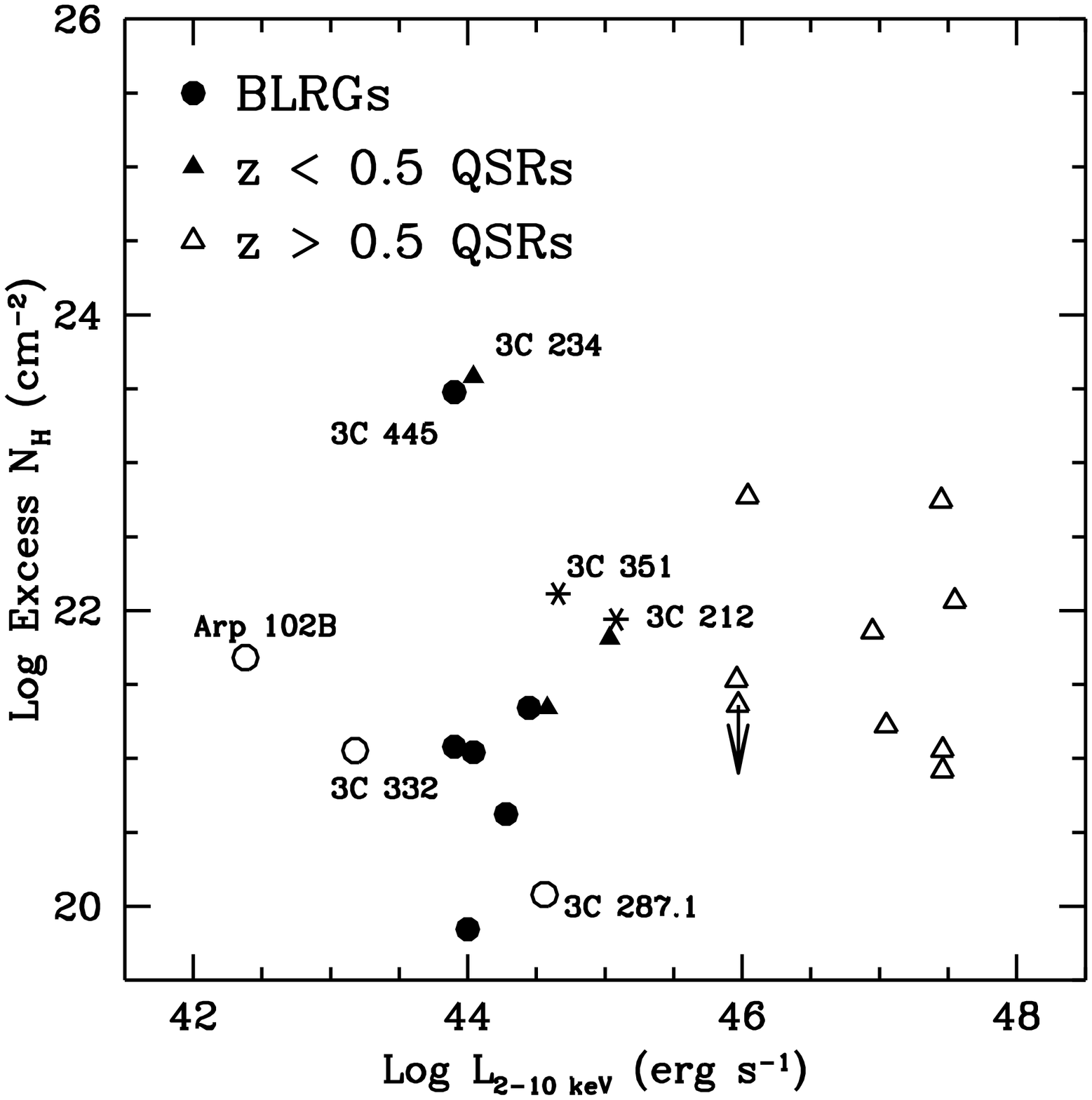,height=3in}}
\caption{{\it (a) Left:} Histogram of the excess X-ray absorption,
defined as the difference between the fitted $N_{\rm H}$ with {\it
ASCA} and the Galactic value, including SIS systematic effects. A
fraction of BLRGs and QSRs, plotted here, have column densities of
cold gas similar to NLRGs and RGs. {\it (b) Right:} Plot of the
intrinsic (rest-frame) excess $N_{\rm H}$ versus luminosity for the
BLRGs and QSRs of our study, together with more distant sources
studied with {\it ROSAT} and {\it ASCA} in the literature. No trends
of the column density with luminosity are apparent over more than five
decades.} \label{fig:nh}
\end{figure}

At soft energies ($<$ 2 keV), we find no evidence for ionized
absorption in the BLRGs of our sample (with the exception of 3C390.3),
in striking contrast to Seyfert 1s. Instead, we detect large columns
of {\it cold} gas.  Figure \ref{fig:nh}a shows the histogram of the
{\it excess} X-ray column density $N{\rm _H^{exc}}$ (defined as the
difference between the fitted X-ray column and the Galactic value in
the direction to the source) for the four classes of RL AGNs. Large
columns, $N{\rm _H^{exc}} \sim 10^{21-24}$ cm$^{-2}$, are measured in
type 2 objects (NLRGs and RGs), presumably due to the obscuring torus
\cite{megunif}. What is more puzzling is the detection of similar
columns in a fraction ($\sim$ 44--60\%) of BLRGs and QSRs, where
according to the unified schemes the line of sight to the nucleus
should be devoid of cold gas.

The excess X-ray columns in the BLRGs and QSRs of our sample are
similar to those observed in more distant RL objects with {\it ASCA} and
{\it ROSAT} \cite{elvis98,cappi95}. This is apparent from Figure
\ref{fig:nh}b, where the intrinsic $N{\rm _H^{exc}}$ (in the source's
rest-frame) is plotted versus the nuclear 2--10 keV luminosity. While
there is a large dispersion in the values of the column density at
lower luminosity, there are no trends over more than five decades: the
more distant sources have similar intrinsic absorbing columns to the
nearby sources.  The origin of the absorbing medium and its location
are not known. However, it is interesting to note that the more
distant objects are mostly core-dominated QSRs, contrary to our
sources, possibly implying an isotropic distribution of the absorber
around the X-ray source.  Future high-resolution observations with
{\it Chandra} and XMM will help clarify the nature of the X-ray
absorber.

\begin{figure}
\noindent
{\psfig{figure=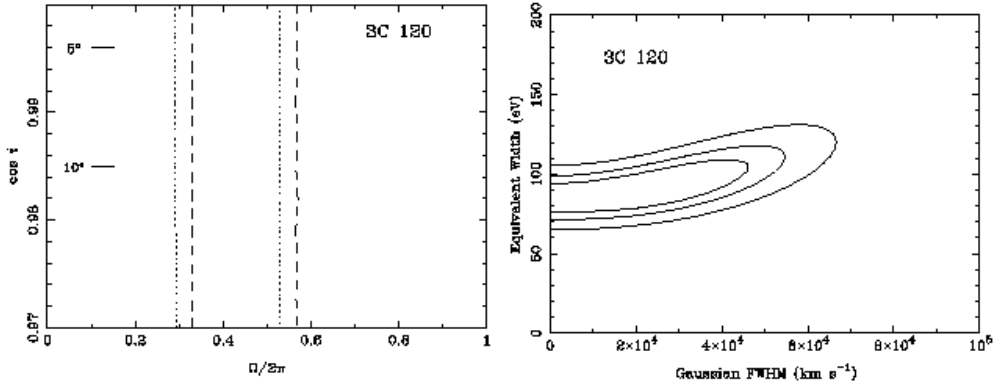,height=2in}} 
\caption{{\it (a) Left:} 68, 90, and 99\% confidence contours in the 
width--equivalent width plane showing the properties of the \feka\
line of 3C~120. The line is obviously unresolved with $FWHM<55,00~{\rm
km~s^{-1}}$, and and an equivalent width of $90^{+30}_{-20}$~eV. {\it
(b) Right:} 90\% confidence contours in the disk inclination
angle--solid angle plane indicating the allowed strength of the
Compton refelection component in 3C~120. The dashed and dotted lines
denote the bounds corresponding to different folding energies of the
primary X-ray spectrum (300 and 600~keV, respectively). These results
are independent of the Fe abundance.}
\label{fig:rxte}
\end{figure}

The \feka~line is detected by {\it ASCA} in 67\% of BLRGs, 20\% of
QSRs, 33\% of NLRGs, and 30\% of RGs. In BLRGs and QSRs the line is
broad, with ${\rm FWHM}$ \gtsima 20,000 km s$^{-1}$, while in NLRGs
and RGs it is unresolved. However, it is difficult to study the line
profile due to the limited sensitivity of {\it ASCA} and the fact the
line is weak in RL AGNs.  Stronger constraints are provided by {\it
RXTE} thanks to its larger collecting area.  The \feka~line is
detected in the BLRGs of the sample (e.g., Figure \ref{fig:rxte}a),
with an Equivalent Width of EW \ltsima 90 eV, lower than what was
measured by {\it ASCA} in most Seyferts \cite{nandra97}.  {\it RXTE}
has been able to detect the \feka\ line in 3C~111 and Pictor~A, where
{\it ASCA} obtained only upper limits of about 100~eV.  Unfortunately,
because of the poor {\it RXTE} resolution, the lines are unresolved
(Figure \ref{fig:rxte}a).

The wide energy range (2--250 keV) covered by the PCA and HEXTE
instruments on {\it RXTE} is well suited to the study of the Compton
reflection in RL AGNs at energies above 10 keV.  The strength of this
component is parameterized in terms of $R=\Omega/2\pi$, i.e., the
fraction of solid angle subtended by the reprocessor to the
illuminating source. We find \cite{ESM99} that in 2/4 sources (3C~111
and Pictor A) a reflection component is not required to model the data
above 10 keV. In 3C~120 and 3C~382, the addition of this component
improves significantly the fit, with strength $R$ \ltsima 0.4--0.5
(see, e.g., Figure \ref{fig:rxte}b). This is lower than what is
observed in most Seyferts \cite{weaver98,lee99}.

\section{Interpretation}

In the currently accepted accretion scenario for RQ AGNs, X-ray
emission is produced in a hot ``corona'' overlaying and illuminating a
standard, Shakura-Sunyaev accretion disk \cite{haardt94}.  In this
picture the reprocessor (disk) subtends a solid angle of $\Omega=2\pi$
to the illuminating source, as observed by {\it Ginga} and {\it RXTE}
in Seyfert galaxies. The \feka~line is emitted within a few
gravitational radii from the black hole \cite{fabian91}, and has a
broad red wing.  Its equivalent width also depends on the solid angle
subtended by the disk to the X-ray source. 

In BLRGs we observe weaker reflection components, indicating $\Omega <
2\pi$. From this perspective, one possible scenario is the ion torus
model \cite{rees82}, recently resuscitated in the form of an Advection
Dominated Accretion Flow (ADAF; \citeasnoun{narayan98}).  Here the
inner parts of the disk (below some transition radius) inflate under
the pressure of the ions which are much hotter than the electrons. The
latter are responsible for the emitted radiation from radio to IR via
synchrotron, bremsstrahlung, and/or inverse Compton emission. The
external cold disk would produce the reflection component and the
\feka~line; both of these features would be weaker than in Seyferts
since the disk subtends a solid angle $\Omega\approx \pi$ to the X-ray
source. While this scenario can also accommodate X-ray slopes similar
to Seyferts, depending on the size of the transition radius between
the inner ADAF and the disk, it does not offer a clear explanation for
the high optical and UV luminosities observed in BLRGs.

An alternative scenario was proposed by \citeasnoun{wozniak98}, on the
basis of similar results to ours obtained from non-simultaneous,
archival {\it ASCA}, {\it CGRO}/OSSE, and {\it Ginga} data. In that
scenario the central engines of BLRGs would be occupied by a standard,
geometrically thin disk but the bulk of the X-ray continuum would be
produced by the inner parts of a relativistic jet.  The beamed X-ray
component would dilute the strength of both the \feka~line and of the
reflection component, the latter produced via Thomson scattering in
the distant molecular torus.  However, our result that RL AGN have
similar X-ray slopes to RQ AGNs argues against a beamed component in
the X-rays. 

In principle, a very good way to discriminate between the two
scenarios outlined above is the Fe line profile. The necessary
observations will be carried out with {\it XMM} and {\it Astro~E}. The
line will have a distinct profile depending on whether it is produced
in the inner regions of a Seyfert-like disk, or at larger distances
(external parts of an ADAF or/and the molecular torus). The current
and future (e.g., {\it Constellation X}) X-ray missions also hold the
potential to measure precisely the spin of the black hole in RL AGNs,
testing models \cite{blandzna77,meier99}, which posit that RL sources
harbor much faster rotating black holes than RQ ones. 

\section{Weak Line Radio Galaxies (WLRGs)}

WLRGs were recently identified as powerful lobe-dominated radio
galaxies with underluminous \oiii~lines and large \oii/\oiii~line
ratios \cite{tadhunter98}, locating these objects in the LINER/H\,
{\sc ii}-region part of the diagnostic line-ratio diagrams of
\citeasnoun{Filippenko96}. A proposed explanation is that these
systems host a low-luminosity AGN \cite{tadhunter98}.

{\it ASCA} observed and detected six WLRGs. In 5/6 cases, the X-ray
spectrum can be decomposed into a hard X-ray component plus a soft
thermal component with $kT \sim 1$ keV. The hard component can be
described by either a flat power law, $\langle \Gamma \rangle=1.5$
(with individual slopes as flat as $\Gamma_X=1.3$), or a very hot ($kT
\sim 100$ keV) thermal bremsstrahlung model. The intrinsic luminosity
of the hard component is $L_{\rm 2-10~keV} \sim
10^{40}-10^{42}$~erg~s$^{-1}$, two orders of magnitude fainter than in
the other radio-loud AGNs of the {\it ASCA} sample. An Fe line with
EW$\sim$ 250 eV is marginally detected in two cases.

\begin{figure}
\noindent
{\psfig{figure=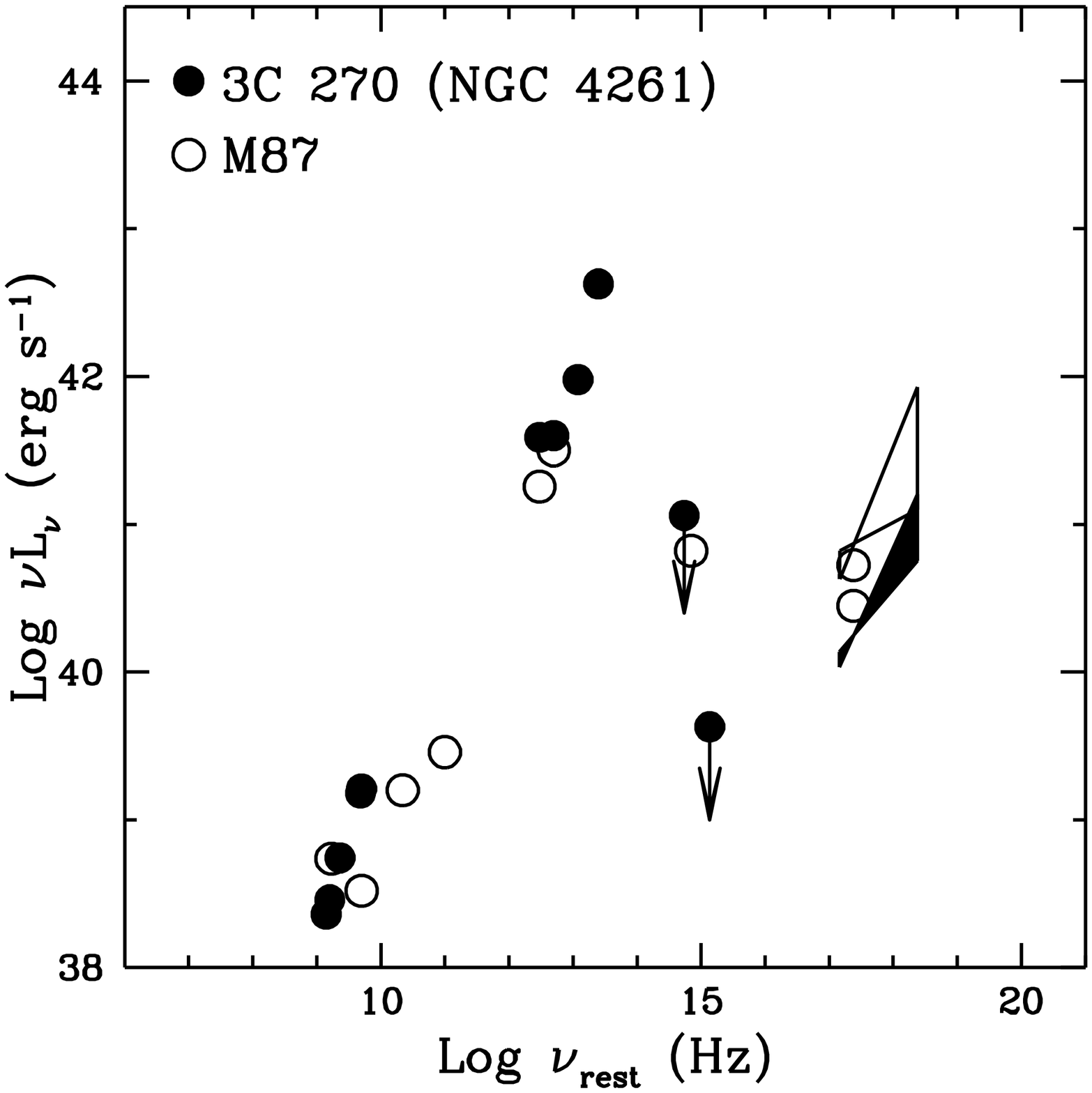,height=2.5in}}{\psfig{figure=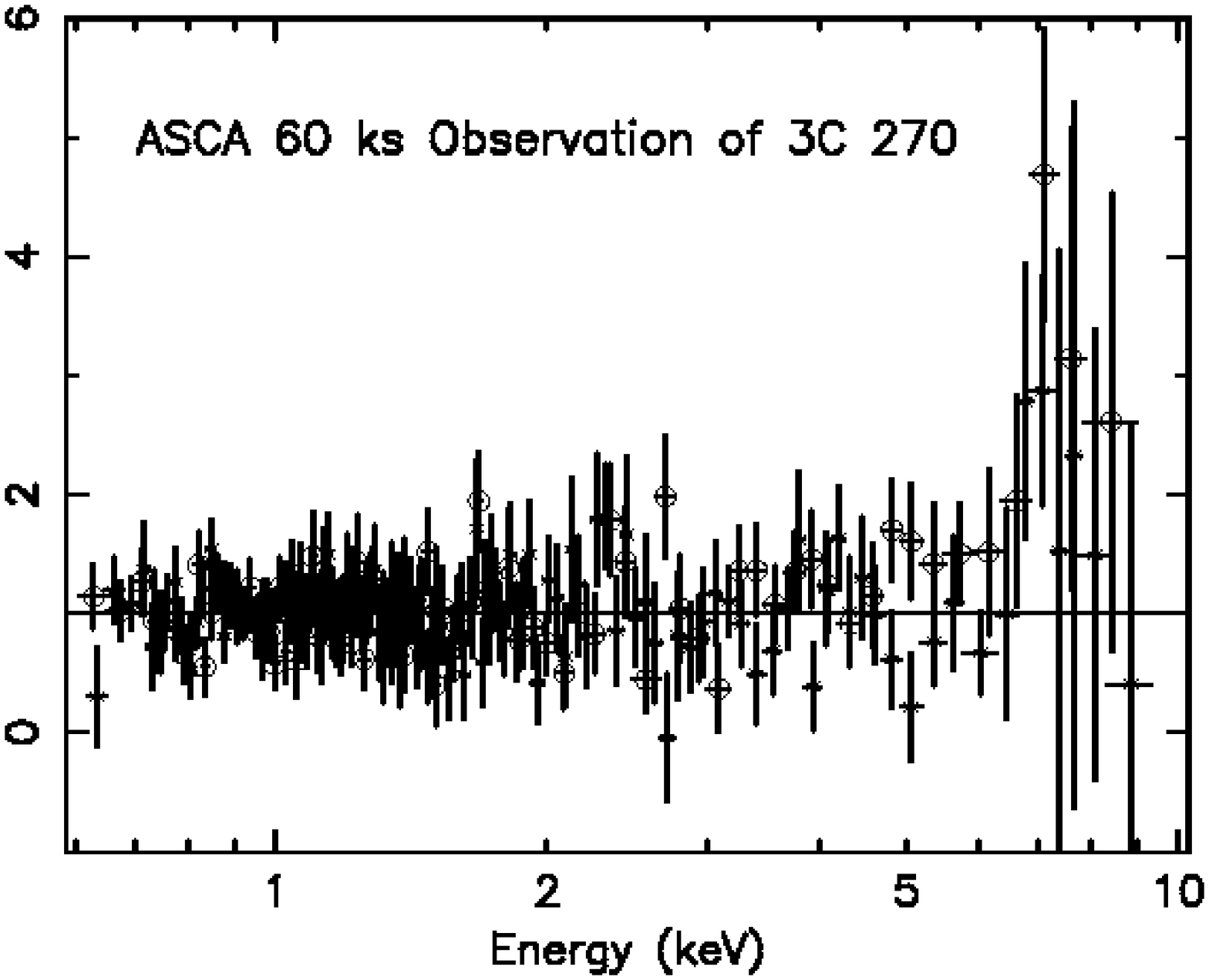,height=3.1in}}
\caption{{\it (a), Left:} Spectral energy distribution of the Weak
Line Radio Galaxy 3C~270 (from {\it ASCA} and literature data),
compared to that of M87. Note the lack of a strong UV continuum. In
both cases, the integrated bolometric luminosity is 10$^{-4}$ of the
Eddington luminosity. {\it (b), Right:} The \feka\ line in the {\it
ASCA} spectrum of 3C~270 (EW $\sim$ 250 eV).  A detailed study of the
line profile, will have to await observations with XMM.}
\label{fig:wlrg}
\end{figure}

To understand in more detail the central engines of WLRGs, we studied
the case of 3C~270, hosted by the nearby giant elliptical NGC~4261,
where {\it HST} observations provide an estimate of the central black
hole mass, $M_{\rm BH}=(4.9\pm 1.0) \times 10^8$ $M_{\odot}$
\cite{ferrarese96}. This implies an Eddington luminosity $L_{\rm Edd}
\sim (5-8) \times 10^{46}$ erg s$^{-1}$. From the radio-to-X-ray
spectral energy distribution of 3C~270 (Figure \ref{fig:wlrg}a), we
estimate a bolometric luminosity of $L_{\rm Bol} \sim 2 \times
10^{43}$ erg s$^{-1}$. Thus, $L_{\rm Bol}/L_{\rm Edd} \sim (2-4)
\times 10^{-4}$, placing 3C~270 in the ADAF regime
\cite{narayan98}. It is thus tempting to speculate that WLRGs
represent that segment of the population of radio-loud AGN in which
the accretion rate is so low that an ADAF is inevitable. In this
context, the lack of a strong far-UV ionizing continuum is the cause
of the observed underluminous \oiii~emission lines.

Interestingly, an Fe line is marginally (90\% confidence) detected in
3C~270 between 6.5--7.5 keV with EW $\sim$ 250 eV (Figure
\ref{fig:wlrg}b).  Future observations of this and other WLRGs at
higher-sensitivity will allow us to study the line profile and energy,
discriminating among possible scenarios for its origin. Particularly
intriguing is the possibility that the Fe line originates in the
coronal gas in the outer parts of the ADAF \cite{narayan99}. In this
case, soft X-ray lines would be expected as well, whose relative
strengths are diagnostics of the size and conditions in the ADAF. Both
the Fe and soft X-ray line emission in WLRGs will be well studied by
{\it Chandra} and XMM, which afford an ideal combination of high
sensitivity and angular resolution.

\section{Conclusions}

We performed an extensive study of the X-ray spectral properties of RL
AGNs using {\it ASCA} and {\it RXTE}. Our results support a picture
where the central engines of RL are systematically different from
their RQ counterparts, with RL AGNs having weaker Fe lines and
reflection components than RQ ones. The differences also extends to
the circumnuclear environments: large columns of cold gas are detected
in a fraction of BLRGs and QSRs, while ionized absorption is more
typical in Seyfert 1s. Our X-ray results suggest that the origin of
the RL/RQ AGN dichotomy must be sought in intrinsic properties of the
central engines of these systems, rather than in their larger scale
environments (host galaxies/clusters).

We reported the first observations at medium-hard X-rays with {\it
ASCA} of a new subclass of radio galaxies, the WLRGs. There is strong
evidence that these systems harbor a low-luminosity AGN in which the
accretion flows are advection dominated (e.g., the case study of
3C~270).

The three major X-ray observatories of the next century, {\it
Chandra}, XMM, and Astro E, will have a central role in allowing us to
study in detail the Fe line profiles, the nature of the mysterious
X-ray absorber, and the structure and geometry of the accretion flow.
Future X-ray observations at high sensitivity and resolution will
provide a giant leap forward in our understanding of RL AGNs
(comparable to the one {\it ASCA} already provided for RQ AGNs),
opening a new perspective on the origin of the RL/RQ AGN dichotomy.

\ack
This work was supported by NASA contract NAS-38252 and NASA grants
NAG5-7276 and NAG5-7733.


\end{document}